# Iron spin crossover in ferropericlase and its effect on lower-mantle thermal conductivity

Alexander F. Goncharov [1]*, Irina Chuvashova [1,2], Eric Edmund [3], Jung-Fu Lin[4], Zena Younes [5], Nicolas Jaisle [5], Axel Phelipeau [6, 7], Carmen Sanchez-Valle [3], Christoph Otzen [7], Clemens Prescher [7], Hanns-Peter Liermann [6], Nico Giordano [6], James McHardy [5], Karen Appel [8], Michal Andrzejewski [8], S. V. Rahul [8], Minxue Tang [8], Jolanta Sztuk-Dambietz [8], Thomas Michelat [8], Torsten Laurus [6], Malcolm McMahon [5], Mark Robertson [5], Rachel Husband [6], Efim Kolesnikov [9], Sebastien Merkel [9], Silvia Boccato [10], Carsten Baehtz [11], Guillaume Morard [10,12], Emma S. Bullock [1], Cornelius Strohm [6], Zuzana Konopkova [8], Ryan Stewart McWilliams [5]*

[1] Earth and Planets Laboratory, Carnegie Institution for Science, Washington, DC, USA

[2] Now at Florida International University, Miami, FL, USA

[3] University of Münster, Institute of Mineralogy Corrensstraße 24, Münster, Germany

[4] Department of Earth and Planetary Sciences, Jackson School of Geosciences, The University of Texas at Austin, Austin, TX, USA

[5] SUPA, School of Physics and Astronomy, and Centre for Science at Extreme Conditions, The University of Edinburgh, Edinburgh, UK

[6] Deutsches Elektronen-Synchrotron (DESY), Notkestr. 85, 22607 Hamburg, Germany

[7] Institute of Earth and Environmental Sciences—Geomaterials and Crystalline Materials, Albert-Ludwigs Universität Freiburg, Hermann-Herder Str. 5, 79104 Freiburg, Germany

[8] European XFEL, Schenefeld, Germany

[9] Univ. Lille, CNRS, INRAE, Centrale Lille, UMR 8207—UMET—Unité Matériaux et Transformations, F-59000 Lille, France

[10] Sorbonne Université, Muséum National d'Histoire Naturelle, UMR CNRS 7590, Institut de Minéralogie, de Physique des Matériaux et de Cosmochimie (IMPMC), Paris, France

[11] Helmholtz-Zentrum Dresden-Rossendorf, 01328 Dresden – Germany

[12] ISTerre, Université Grenoble Alpes, CNRS, Grenoble, France

* Corresponding authors

Thermal conductivity of Earth's lower mantle controls heat transfer across the core–mantle boundary (CMB) and strongly influences mantle convection. We report direct measurements of the thermal conductivity of single-crystal ferropericlase ($Mg_{(1-x)}Fe_xO$, x = 0.09–0.13), the second most abundant lower-mantle mineral, using optical laser flash and X-ray free-electron laser heating in diamond-anvil cells up to ~2200 K and 130 GPa. These experiments provide the first conductivity data for ferropericlase at simultaneous lower-mantle pressures and temperatures. A marked reduction in conductivity between 60–100 GPa at ~1700 K is consistent with the iron spin crossover. Combined with our previous results for Fe- and Fe,Al-bearing bridgmanite, the data define a lower-mantle conductivity profile increasing with pressure to ~10 W·m$^{-1}$·K$^{-1}$ near the CMB, constraining mantle heat flux, plume buoyancy, and long-term geodynamic evolution.

## 1. Introduction

Heat transport in the Earth's interior governs the planet's long-term thermal and dynamical evolution. In the lower mantle, extending from ~660 to 2900 km depth, heat is transferred both by thermal conduction and by convection of silicate and oxide materials. The efficiency of conductive heat transport controls the heat flux across the core–mantle boundary (CMB) into the lower mantle, influencing major geodynamic processes, including the maintenance of the geodynamo, the formation and longevity of mantle plumes, and Earth's global thermal evolution [1-6]. Accurate knowledge of the thermal conductivity (k) of lower-mantle minerals under appropriate pressure–temperature (P–T) conditions is therefore essential for quantifying the planet's internal heat budget. In particular, the balance between conductive and convective heat transport regulates mantle dynamics and the development of large-scale structures. A detailed description of the temperature and depth dependence of thermal conductivity is thus crucial for understanding mantle geochemical heterogeneity, constraining the stability of plumes, and elucidating the coupling between mantle and core evolution [7-10].

The lower mantle is composed primarily of bridgmanite (Mg,Fe)SiO$_3$ (Bgm), which transforms to post-perovskite (Ppv) in thin layers in the lowermost mantle [11]), ferropericlase (Mg,Fe)O (FP), and CaSiO$_3$ perovskite, with FP constituting up to ~20% of the volume (e.g., Ref. [12]). Although less abundant than bridgmanite, FP may exert a disproportionately strong influence on bulk mantle thermal properties owing to its relatively high intrinsic thermal conductivity [13-16].

A key feature of FP is the pressure-induced electronic spin crossover of Fe$^{2+}$ from the high-spin (HS, S = 2) to the low-spin (LS, S = 0) state between ~50 and 70 GPa at room temperature [17,18]. This transition modifies the electronic, elastic, vibrational, and structural properties of the lattice and may significantly affect mantle thermal transport [19-24].

Where conduction dominates heat transport near the CMB, under conditions where the iron spin crossover is essentially complete [25,26], the contribution of highly conductive FP together with its spatial distribution, volume fraction, and composition can strongly influence the aggregate thermal conductivity, thereby controlling the local and global heat flux across the CMB from the core into plume-forming regions. Accurately quantifying the contribution of FP is therefore essential for constraining the thermal evolution of the whole Earth system. Although convection dominates heat transport in the mantle, thermal conductivity controls the background temperature gradient and viscosity structure, and, therefore can influence mantle dynamics throughout the lower mantle, not only in the thermal boundary layer (TBL) near the CMB [15,26,27].

Previous measurements of thermal conductivity in FP have been largely restricted to ambient-temperature or low-pressure conditions [13,16]. The limited high-pressure studies that do exist were performed at room temperature, far from the high-temperature regime of the lower-mantle geotherm [15,28]. Although theoretical studies predict strong effects of the spin crossover on the thermal conductivity of $(Mg,Fe)O$ [27,29], direct experimental constraints at simultaneous lower-mantle pressure–temperature conditions remain sparse and inconsistent [13,15,16,28,30], and further are complicated by the use of powder samples, which introduce extrinsic scattering mechanisms [28]. As a result, the magnitude of FP thermal conductivity and the effect of the iron spin crossover on FP thermal conductivity remain debated even at room temperature and essentially unconstrained at the elevated temperatures relevant to the deep mantle. Consequently, the behavior of FP thermal conductivity under the combined high-pressure and high-temperature (P–T) conditions of the lower mantle including those approaching the thermal boundary layer (TBL) remains poorly constrained by experimental data.

Here, we report direct measurements of the thermal conductivity of FP ($Mg_{(1-x)}Fe_xO$, x = 0.09–0.13; FP9–13) under simultaneous high-pressure–high-temperature conditions using optical laser flash heating and X-ray free-electron laser (XFEL) heating in diamond anvil cells (DACs). Our experiments span pressures up to 130 GPa and temperatures up to 2200 K, approaching those of the lower mantle geotherm. The results reveal a pronounced reduction in thermal conductivity across the iron spin crossover, followed by a recovery at higher pressures. When combined with our previous measurements on Fe- and Fe,Al-bearing bridgmanite [31], these data constrain the aggregate thermal conductivity profile of the lower mantle, demonstrating a substantial increase toward the CMB and offering new insights into mantle convection, plume buoyancy, and Earth's global heat budget.

## Results and discussion

### XFEL heating

The experiments were conducted at three pressure conditions using three DACs, with FP samples precompressed to initial pressures of 50, 74, and 120 GPa. The samples were heated up to 2800 K. The P–T paths at 50 and 74 GPa crossed the high-spin (HS) to mixed-spin (MS) and MS to low-spin (LS) boundaries as defined in Ref. [25] (Figure S2, *Supplementary Materials*). Here, we assume that phase lines of our FP10 are close to those measured by [25] for their FP19, while the spin transition for FP25 probed in Ref. [26] is shifted to higher pressures compared to those for FP19. Under this assumption, the high-temperature run at 50 GPa sampled the HS state predominantly, whereas the 74 GPa run mainly probed the MS state. The 120 GPa experiment accessed exclusively the LS state.

Samples were heated at 2–4 different locations, gradually increasing the XFEL transmission at each position. XRD patterns of the heated samples show peak shifts to lower angles and peak broadening (Figure S4, *Supplementary Materials*), reflecting thermal expansion and temperature gradients. These observations are consistent with a temperature distribution having a maximum near the center of the cavity, as predicted by finite-element (FE) calculations [31]. In this configuration, the XRD peak shape is dominated by the contribution from the hottest, thermally expanded material near the cavity center. Only runs that produced substantial peak shifts and thus measurable thermal expansion were selected for further analysis.

In experiments where the temperature exceeded ~2000 K, the maximum sample temperature could also be determined by spectroradiometry. These radiative temperatures are in good agreement with those obtained from FE calculations based on the thermal expansion of the central region of the sample cavity (Figures S2, S3, Supplementary Materials). This agreement supports the assumptions about thermal pressure used in our analysis. Remaining discrepancies may arise from deviations from the graybody approximation in radiative temperature measurements [32]. In any case, the inferred temperatures depend only weakly on the assumed thermal pressure and do not affect the determination of thermal conductivity.

Runs reaching significantly higher temperatures (~2500 K and above, as estimated from spectroradiometry) typically resulted in peak splitting. These runs were discarded, and fresh positions were selected for subsequent experiments. The heated regions often showed visible changes in the form of dark spots, likely caused by incongruent melting and chemical segregation accompanied by Fe migration [33].

The XRD data capture the time evolution of three Bragg reflections (111), (200), and (220) during heating (Figure S4, *Supplementary Materials*). These reflections were used to determine instantaneous sample volumes by least-squares refinement of the unit-cell parameters. At 50 GPa, only the (200) reflection was reliably observed, and the lattice parameter was therefore derived from this peak alone. Comparisons at higher pressures show that the lattice parameter obtained from different Bragg reflections differ by less than 3%, indicating that single-peak determination at 51 GPa provides sufficient accuracy.

The temperature histories derived from in situ XRD measurements of the sample unit-cell volume show an abrupt thermal expansion at the beginning of each run (Figure 1), followed by gradual leveling off and slow contraction toward the end. The initial expansion reflects heat accumulation in the sample due to absorption of high-repetition-rate XFEL pulses. The XFEL pulse envelope typically decreases slowly with time (Figure S1, *Supplementary Materials*), leading to gradual cooling of the sample at the end of the run. In the finite-element (FE) calculations, variations in pulse intensity within the train were explicitly taken into account, and the resulting temperature changes after reaching steady state were used to constrain the temperature dependence of thermal conductivity, following the approach developed in our previous work on solid argon at comparable time scales [34].

The temperature histories are well reproduced by our FE calculations when using appropriately chosen thermal conductivity values (Figure 1). Initially, each experimental curve was analyzed assuming temperature-independent thermal conductivity. Various temperature dependencies of thermal conductivity were then tested to improve agreement with the experimental data, and in some cases, the fit was significantly enhanced. The resulting temperature-dependent thermal conductivity values are presented in Table S1 in the *Supplementary Materials*, along with the thermal conductivity values at 2000 K, which closely correspond to the P–T conditions of the lower-mantle geotherm (Figure S2, *Supplementary Materials*). The resulting pressure dependence of thermal conductivity is nonmonotonic, in agreement with observations from the laser flash heating experiments described in the next section.

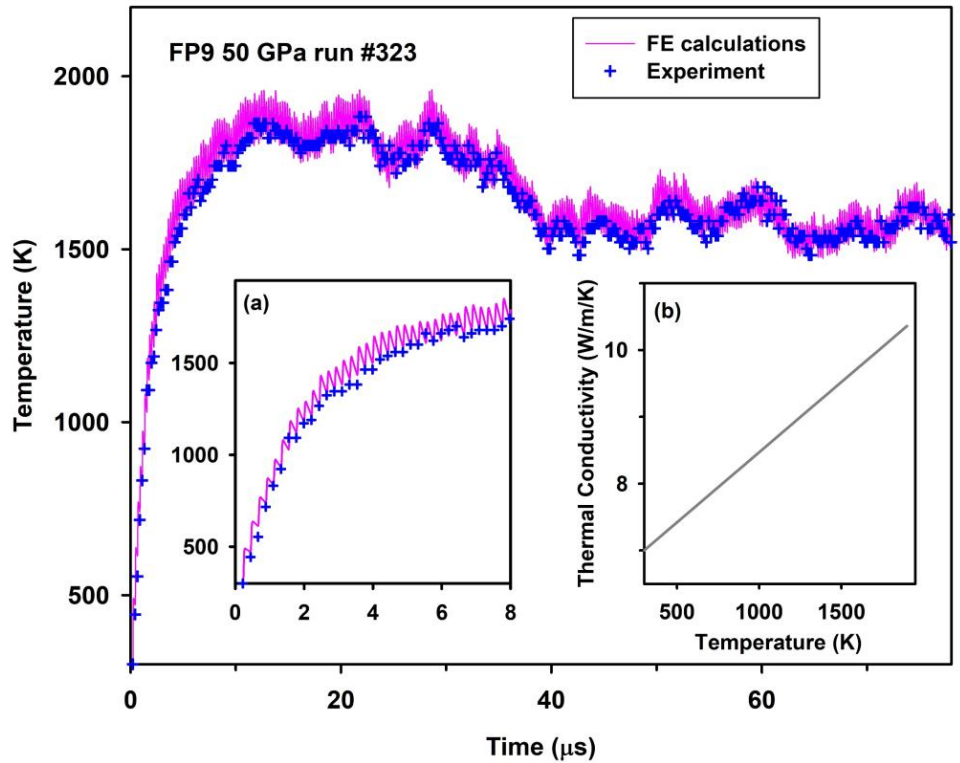

**Figure 1.** Heating history of FP in the DAC during EuXFEL proposal 8025, run 323, experiment CC278. Crosses represent experimental data corresponding to a heating run at a nominal pressure of 50 GPa. Inset (a) shows an expanded view of the initial stage of XFEL heating. Temperatures during XFEL heating were determined from the thermal expansion measured in XRD experiments, using the experimental thermal equation of state [25] and assuming 30% of the isochoric thermal pressure [35]. The experimental data are compared with FE calculations at the cavity center (lines) assuming temperature-dependent thermal conductivity (inset (b)). The FE models also incorporate temperature-dependent sample density [25] and theoretically calculated heat capacity [36], and use the experimentally measured XFEL pulse energies.

**Laser flash heating**

The experiments were conducted using several DACs, each targeting a specific pressure range. Typically, a single DAC sample was measured at 2–4 pressure points (Table S2, *Supplementary Materials*). Pressure was determined from the Raman shift of the diamond edge [37]. The samples used in these experiments were of composition FP13.

In these experiments, the sample was quasi-continuously preheated from both sides to 1400–2200 K depending on the run, and a 2 μs nearly rectangular laser pulse was applied to one side, producing a rapid temperature rise. The temperature modulation on the heated side was set to 200–300 K. The temperature history recorded on the opposite (probe) side of the sample exhibits a delayed and temporally broadened response relative to the heated side as well as a lower amplitude (Figure 2). The temperature histories recorded from both sample sides were fitted using FE calculations, in which the thermal conductivity of the FP sample and KCl medium were treated as the free parameters (Table S2 in the *Supplementary Materials*).

The data from the in-house laser flash heating experiments are shown in Figure 2 for the case of 125 GPa. Because of the relatively low thermal conductivity of FP, the temperature variation on the probe side of a few-micrometer-thick sample is much smaller than that on the pulsed-heated side. Consequently, the temperature history measured on the probe side becomes sensitive to experimental artifacts, such as direct heating by a small fraction of the pulsed laser (approximately 2% of the incident power) transmitted through imperfections in the polarization optics. This effect was explicitly incorporated into our finite-element (FE) model when calculating the probe-side temperature history, as illustrated by measurements performed at 68 GPa (Figure S6, *Supplementary Materials*).

The measured thermal conductivity corresponds to a relatively narrow temperature interval, from the preheating temperature to the maximum reached during pulsed heating. Consequently, its temperature dependence cannot be reliably constrained and was approximated as constant over this range (typically 1700–2400 K), as reported in Table S2 in the *Supplementary Materials*.

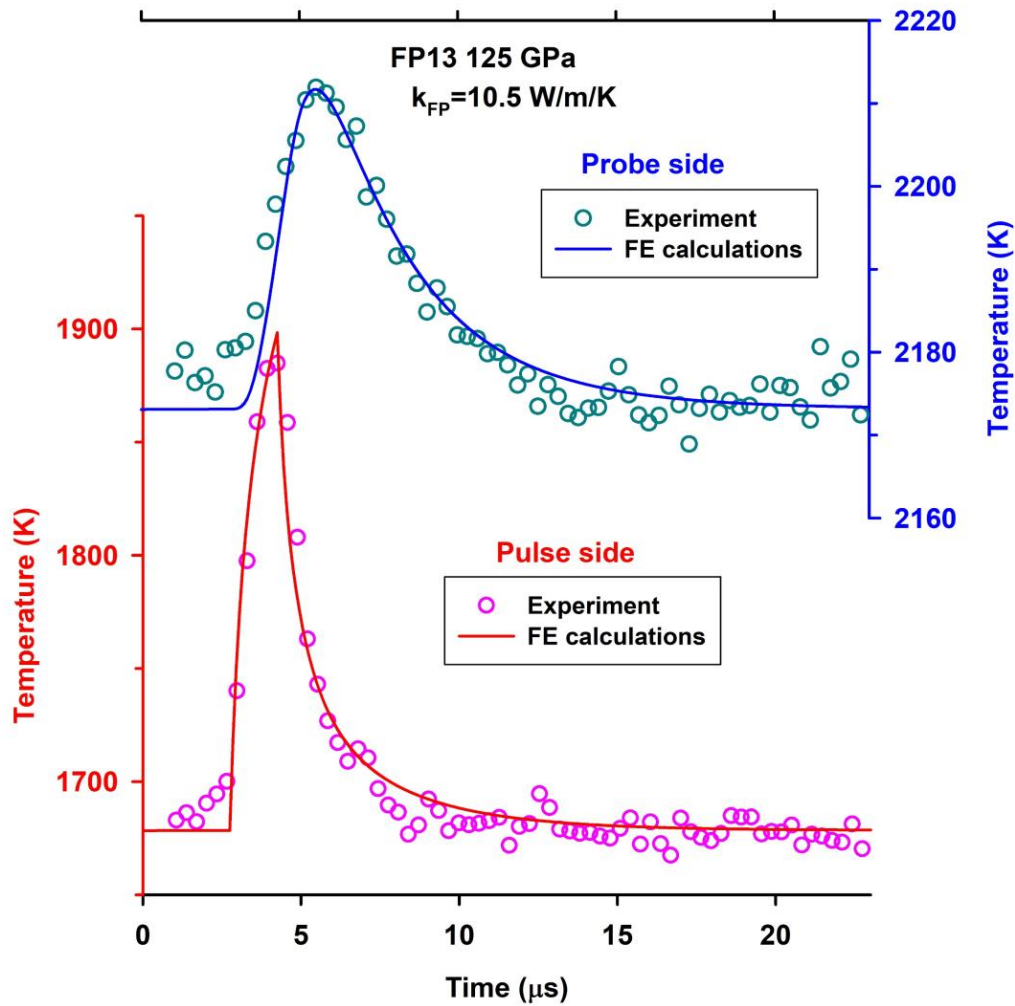

**Figure 2.** Heating histories in an in-house laser flash heating experiment on FP13 at 125 GPa. Open circles show experimentally measured radiative temperatures, with pink and cyan corresponding to the pulsed-heated and probe sides of the sample, respectively. Solid lines represent the best fits obtained from FE calculations [38,39]. The parameters used in the fit are listed in Table S2 in the *Supplementary Materials*.

**Pressure-dependent thermal conductivity at geotherm**

The pressure dependence of thermal conductivity is shown in Figure 3, with emphasis on the values at 2000 K, which closely represent the lower mantle geotherm (Figure S2, *Supplementary Materials*). The experiments reported here were performed at temperatures near this value, so only minor corrections if any, as in the case of the flash heating experiments, were needed to determine thermal conductivity at 2000 K (Table S1, *Supplementary Materials*).

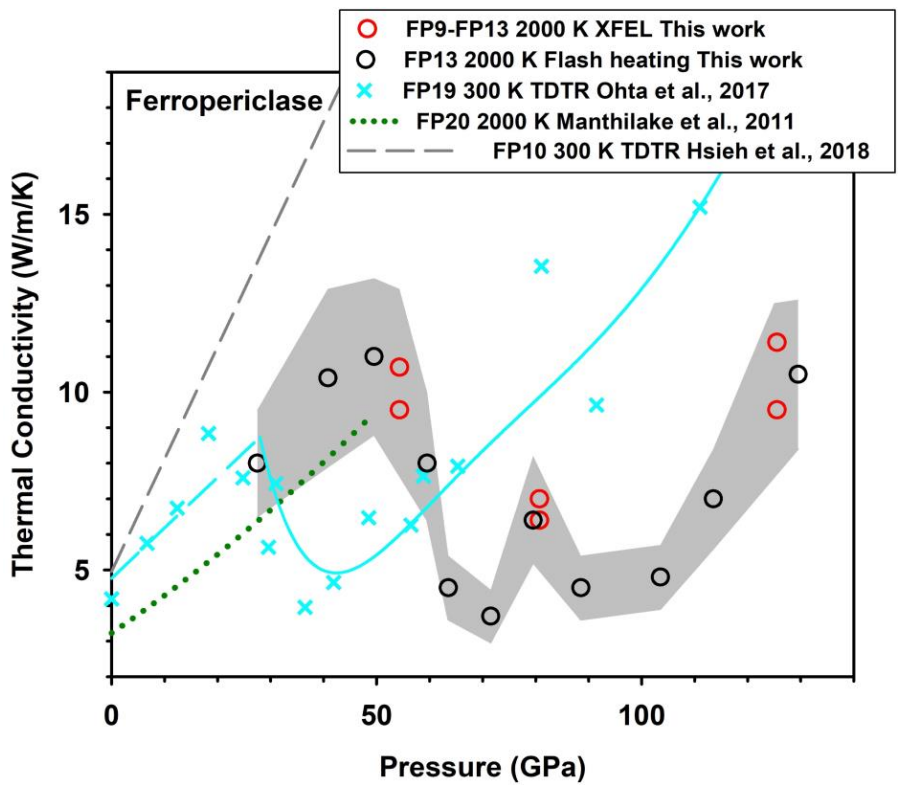

**Figure 3.** Thermal conductivity of FP as a function of pressure at high and room temperatures. Circles show results from this work, obtained from XFEL and in-house laser flash heating experiments, with the gray shaded area indicating the measurement confidence interval, determined using the error propagation of the uncertainties in the parameters (Tables S1 and S2, *Supplementary Materials*) used to deduce the thermal conductivity values. Thin cyan crosses represent FP19 at 300 K from Ref. [28], and the long-dashed gray line shows FP10 at 300 K data from Ref. [15]. The dotted dark green line is the extrapolation of large-press Ångström method measurements for FP20 at 1273 K and 14 GPa [16] to higher pressures at 2000 K. These results are in close agreement with additional pulsed-heating data from large-press apparatus experiments reported in Ref. [13] up to 15 GPa and 873 K. Representative Liebfried Schloemann model [40] fits to our data can be found in Supplementary Figure S7.

Our results reveal a nonmonotonic pressure dependence of thermal conductivity, characterized by a sharp drop above 50 GPa, a broad plateau with a weak maximum between 60 and 105 GPa, and a rapid increase at higher pressures. This behavior can be naturally associated with the iron spin transition, as the thermal conductivity of mixed HS–LS phases is expected to decrease [27,29], consistent with anomalous softening of the elastic moduli in the transition region [19-24]. Notably, a small peak in thermal conductivity near 80 GPa was predicted theoretically by [27] and attributed to partial cancellation of competing factors controlling heat transport. In particular, the heat capacity exhibits anomalous behavior across the spin transition [36], which can produce a positive temperature dependence of thermal conductivity in this region, consistent with our XFEL measurements within the spin-transition interval (Fig. 1). To our knowledge, these experiments provide the first direct experimental evidence for the characteristic two-valley pressure dependence of FP thermal conductivity, confirming the anomalous transport behavior associated with the spin transition.

The magnitude of the spin transition effect is substantial: we observe a drop in thermal conductivity of more than 50%, compared to less than 20% predicted by [27]. However, the observed effect is comparable to that reported by [28] at 300 K. In their study, the anomaly occurred at lower pressures, which is broadly consistent with the expected shift of the spin transition with temperature. Overall, their thermal conductivity values for FP19 are significantly lower than ours, which can be attributed to larger Fe content. In addition, their polycrystalline sample, which could add additional thermal resistance due to the grain boundaries. Also, their data were obtained at near-room temperature, whereas our data mainly refer to high temperatures (ca. 2000 K).

Interestingly, another 300 K study on FP10 by [15] disagrees both with [28] and with our results, as no signature of the spin transition was detected in that work. This discrepancy is surprising, particularly because similar types of samples were used in both studies. Notably, the pressure dependence of our thermal conductivity below the spin transition agrees well with P–T extrapolations from low-pressure data reported by reported by [16](<14 GPa, <1300 K) and [13](<15 GPa, <900 K).

**Implication for the heat transport in the lower mantle**

We applied the new thermal conductivity data obtained at extreme P–T conditions to develop a model for the lower-mantle thermal conductivity (Fig. 4). The lower mantle was assumed to consist of 20% FP and 80% Bgm, as commonly accepted. For the P–T-dependent thermal conductivity of Bgm, we used results from recent work employing similar experimental techniques [31]. Contributions from radiative thermal conductivity were neglected [41]. The spin transition in FP has a moderate effect on the lower-mantle thermal conductivity between 1450 and 2450 km depth, producing a dip of approximately 10%. This effect can be enhanced in the lower mantle due to the increased connectivity at these P-T conditions [42].

This contrasts with previous models [16,43], which were largely dependent on extrapolations to high P-T conditions at the geotherm.

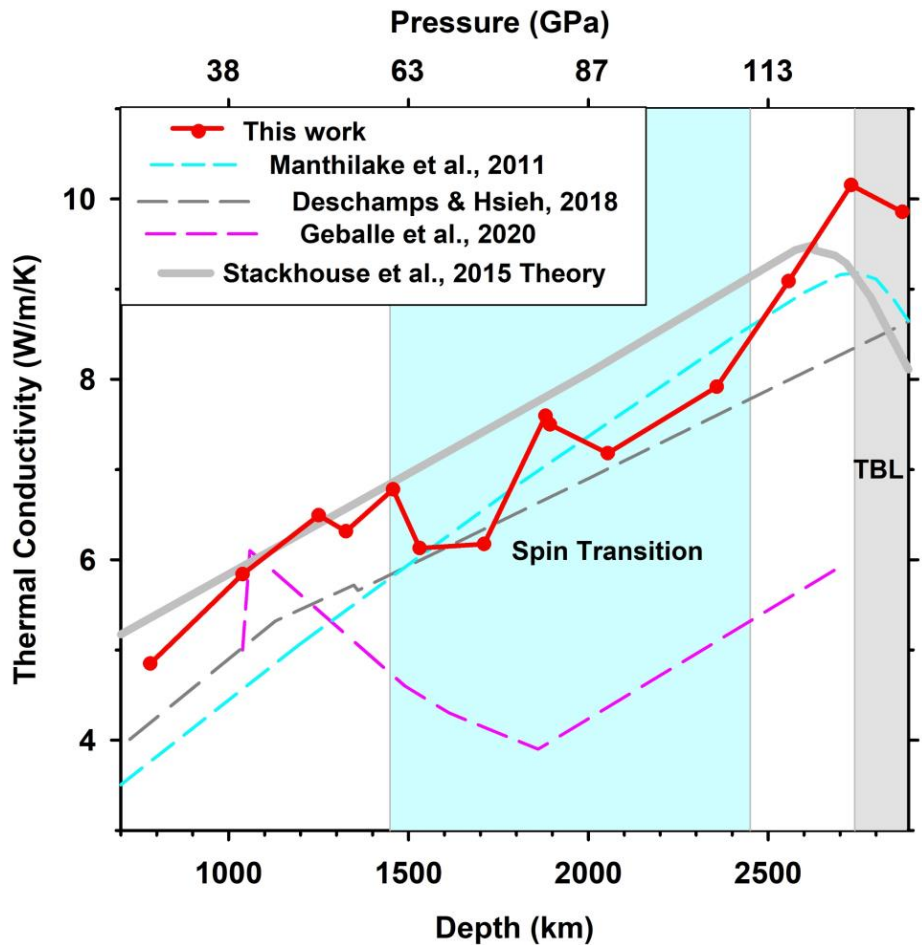

**Figure 4.** Thermal conductivity models of the lower mantle computed from our results using Hashin–Shtrikman averaging [44] for aggregates of 20% FP + 80% Bgm at P–T conditions along the geotherm. FP data correspond to $Mg_{(1-x)}Fe_xO$ (x = 0.09–0.13) from this work; Bgm data for Fe- and Fe,Al-bearing compositions are from experiments using the same techniques [31]. For a smoothened version of this dependence please see Figure S7 in *Supplementary Materials*. These results are compared with models based on Ångström measurements in a large press up to 26 GPa and 1273 K [16], TDTR measurements at 300 K up to 120 GPa [43] extrapolated to CMB P–T conditions, and direct laser flash-heating experiments of pyrolite aggregates up to 120 GPa and 2500 K [45]. Ab initio theoretical calculations for the pyrolite mantle [46] are also shown. Shaded areas indicate the depth ranges of the FP spin transition and the thermal boundary layer.

In the depth range corresponding to the mid-mantle spin transition resolved in our data, the reduced thermal conductivity suggests the presence of a potential dynamical transition zone within the mid-lower mantle. Such variations may promote more sluggish convection in the overlying layers [47] and contribute to regional variability in the Rayleigh number. Overall, our model predicts slightly higher thermal conductivity values than most previous studies, although the differences are generally within ~20%, comparable to the experimental uncertainty. The results also agree well with *ab initio* calculations [46], except within the pressure range of the FP spin transition, where our measurements indicate lower thermal conductivity (Figure 4).

Our results are qualitatively consistent with laser-flash heating experiments on mineral aggregates of pyrolitic composition [45], which likewise show a reduction in thermal conductivity over a broad depth interval (Fig. 4), although the absolute values differ. We tentatively attribute this discrepancy to the ultra-dispersed nature (very small crystallite size) of the samples used in the DAC experiments of[45], where the assumptions underlying Hashin–Shtrikman averaging are unlikely to be strictly valid.

The estimated thermal conductivity of the lowermost mantle near the CMB is approximately 9.8(3.0) $W·m^{-1}·K^{-1}$, in close agreement with previous reliable estimates [16,43,46]. Importantly, our values are based on direct measurements at extreme P–T conditions, requiring only minimal extrapolation to geotherm conditions. These results correspond to a CMB heat flux of 14(4) TW (calculated following Ref. [48]), consistent with the most recent geophysical estimates [49].

**Conclusions**

Our measurements provide the first direct determinations of the thermal conductivity of ferropericlase ($Mg_{(1-x)}Fe_xO$, x = 0.09–0.13) under simultaneous high-pressure and high-temperature conditions characteristic of the lower mantle. Using optical laser-flash heating and XFEL-based pulsed heating in diamond anvil cells, we obtained data up to ~2200 K and 100 GPa. The results show a pronounced decrease in thermal conductivity between 60 and 100 GPa at ~1700 K, consistent with the iron spin crossover from high-spin to low-spin states. When combined with our previous measurements on Fe- and Fe,Al-bearing bridgmanite, these data define a consistent thermal conductivity profile for the lower mantle, increasing strongly with depth reaching ~10 $W·m^{-1}·K^{-1}$ near the core–mantle boundary. This revised profile provides improved constraints on heat flow across the CMB, the stability of deep-mantle plumes, and the geodynamic processes governing mantle convection and plate tectonics.

## Methods

### Sample preparation

Single-crystal FP samples with nominal compositions $(Mg_{(1-x)}Fe_x)O$ (x = 0.09–0.13) were synthesized via inter-diffusion of Fe and Mg between a single-crystal periclase and pre-synthesized (Mg,Fe)O powder in a $H_2/CO_2$ gas-mixing furnace at the Institute for Study of the Earth's Interior (ISEI) of Okayama University at Misasa as reported in Ref. [50]. The MgO crystal with a pre-oriented (100) crystallographic plane purchased from the MTI Corporation was cut to 7 mm in length, 7 mm wide, and 0.25 mm thick. It was sandwiched between two layers of compacted polycrystalline $(Mg_{0.75}Fe_{0.25})O$ powder of approximately 1 mm thick each. The starting sample assemblage was then placed in a Pt holder in the furnace, operating at 1350 °C and an oxygen fugacity of $10^{-2}$ Pa, for approximately 2 weeks. Under these conditions, the $Fe^{3+}$ content in the samples is expected to be very low, consistent with the high optical clarity of the crystals (slightly pale yellow), as previously reported (e.g. Ref. [51,52]). The synthesized single-crystal FP was then extracted and polished for further sample analyses. The recovered pellets were analyzed for chemical homogeneity using electron microprobe analysis (EPMA) and phase purity using X-ray diffraction (XRD). The synthesized materials were confirmed to be single-phase FP with the rock-salt ($Fm\bar{3}m$) structure. Thin single-crystal sections (6-10 μm thick, 20–30 μm in diameter) were then prepared by mechanical double-side polishing to achieve uniform thickness and parallel surfaces suitable for laser heating and XFEL heating experiments.

### High-pressure generation

High pressures up to ~130 GPa were generated using symmetric DACs equipped with diamond anvils with culet diameters of 200 μm and beveled anvils with 100 μm culets. Rhenium gaskets were pre-indented to ~10–25 μm thickness and laser-drilled to produce sample chambers 50–80 μm in diameter.

For in-house optical flash-heating experiments, the FP sample plates were sandwiched between two ~1 μm-thick iridium foils, which served as both laser absorbers and emitters. These multilayer samples were insulated from the diamond anvils by thin KCl slabs to minimize heat loss. In the XFEL experiments, targeting the critical pressure range of 50 and 74 GPa, single-component sample experiments were made possible by loading bare sample plates that precisely bridged the diamond anvils, with argon (gas-loaded) used to fill residual space around the plate in the sample cavity. This approach reduced inhomogeneous deformation and eliminated crushing and cracking, with single-crystal like diffraction patterns of compressed samples validating preservation of a nearly single-crystalline sample. Pressure values were estimated using the Raman shift of the diamond edge, and for XFEL experiments, also by *in situ* X-ray diffraction measurements of the sample density at PETRA III and the European XFEL (EuXFEL), from which pressure was derived using the equation of state of FP [25].

**Laser flash and XFEL heating**

Thermal conductivity was measured using both pulsed laser flash heating [39,53] and X-ray free-electron laser (XFEL) heating techniques [31].

In the laser flash heating measurements conducted at the Carnegie Institution for Science, the sample was continuously heated from both sides and additionally subjected to pulsed heating from one side to generate a heat wave across the sample. Thin iridium (Ir) foils on both sides served as laser absorbers and radiative temperature transducers, enabling measurement of the temperature response used to determine the heat-wave propagation speed—and thus the thermal conductivity. Short-duration infrared laser pulses (2 µs), shaped by a Pockels cell to produce an approximately rectangular profile, greatly simplified the analysis of the transient response. The resulting temperature evolution on each side of the sample was recorded using a streak camera, providing temperature–time histories from which the thermal conductivity was extracted. The thicknesses of the sample and the Ir heat transducer were determined *in situ* at high pressures by optical interferometry, using the optical refractive index of the KCl pressure-transmitting medium scaled to high pressure using the Gladstone-Dale relation [31].

XFEL heating experiments were performed at the High-Energy Density (HED) instrument of the EuXFEL in Schenefeld during User Beamtime 8025 (March 2025) [54]. The X-ray pulses had an energy of 18.1 keV and a duration of <50 fs. The intra-train repetition rate was set to 4.54 MHz, with each pulse train comprising 352 pulses separated by 221 ns. The X-ray beam was focused to a spot size of 5.5 µm (FWHM) in diameter on the sample. Thermal conductivity measurements at EuXFEL rely on the sample's ability to volumetrically absorb a high-repetition train of X-ray pulses [55]. The absorbed energy accumulates in the material, causing its temperature to increase in a stepwise manner with each successive pulse until reaching a steady state. Between pulses, natural cooling of the sample is weak compared to the cumulative heating, which dominates the temperature evolution. Time-resolved X-ray diffraction measurements probe the structural state after cooling following each pulse in the train, allowing the thermal evolution of the sample to be tracked and used to determine the thermal conductivity.

The XFEL train energy was adjusted using absorption filters, with individual pulse energies within each train balanced to within <20% variation (Figure S1 of the *Supplementary Materials*). Per-pulse energies, measured by multiple photodetectors [55], were incorporated into the finite element (FE) simulations either through polynomial fits to their intensity profiles or directly using the raw tabulated pulse-energy data.

The deposited energy causes the sample temperature to rise sequentially with each pulse, as most of the accumulated heat cannot diffuse out of the system between pulses due to the much faster pulse repetition rate relative to thermal transport. The sample temperature, therefore, increases stepwise until a steady state is reached. In the finite element (FE) calculations, the thermal conductivity is determined by the rate at which the system approaches this steady state [31] and by the evolution of the sample temperature under conditions of relatively slow variations in pulse energy across the pulse train.

Pressure at room temperature was determined from the Raman shift of the stressed diamond anvils [37]. Subsequent XRD measurements of FP samples conducted at the Extreme Condition Beamtime at PETRA III [56], DESY (Hamburg) and the HED instrument at EuXFEL yielded consistent pressure values when evaluated using the equation of state [25].

Temperature in these experiments performed at various pressures (50-120 GPa, Figure S2, *Supplementary Materials*) was determined from the thermal expansion measured in time-domain XRD experiments using the thermal equation of state of FP [*Komabayashi et al.*, 2010], assuming that the thermal pressure constitutes 30% of the maximum thermodynamic isochoric pressure, $P_{th} \cong 0.3\alpha K_T \Delta T$, where $\alpha$ and $K_T$ are the thermal expansivity and isothermal bulk modulus, and $\Delta T$ is the temperature increase during heating [35]. This approximation was also employed in our previous work [31], where it yielded internally consistent results. In a recent XFEL laser-heating study [57] an independent attempt was made to directly determine the thermal pressure, resulting in somewhat different pressure-dependent values of $P_{\text{th}}$. The temperatures obtained from these calculations were in qualitative agreement with the streak spectroradiometry measurements performed *in situ* during the heating runs (Figures S2 and S3, *Supplementary Materials*).

Thermal conductivity values were extracted from finite element (FE) model calculations (see details in Ref. [31]) in which time-dependent temperature maps of the high-pressure cavity were computed in a 2D approximation using experimentally determined in situ at high pressures geometries (300 K) and densities (at all probed temperatures), along with literature thermodynamic parameters [25,58]. The sample thickness was determined *in situ* by optical interferometry using the experimentally measured optical refractive index of MgO [59]. The sample thickness is not a critical parameter in the determination of thermal conductivity using the XFEL heating technique [31]. Therefore, any uncertainty associated with the difference in refractive index between MgO and FP can be neglected. Based on a linear interpolation between MgO and FeO, the refractive index of $Mg_{0.9}Fe_{0.1}O$ (FP10) is expected to be approximately 3% higher than that of MgO, which would have a negligible effect on the derived thermal conductivity.

**Post-experimental characterization**

Recovered samples were examined using scanning electron microscopy (SEM) and electron microprobe analysis (EPMA) to assess chemical homogeneity, possible reactions with the pressure medium, and overall structural integrity. X-ray diffraction measurements on quenched samples confirmed the preservation of the rock-salt structure and showed no evidence of decomposition or phase segregation.

In flash heating experiments, the sample and iridium transducer thicknesses were confirmed using FIB measurements on the recovered sample.

Scanning Electron Microscopy (SEM) images of samples recovered after the XFEL experiments show that the material remained chemically uniform upon recovery, as confirmed by multiple measurements at different points (Figure S4, *Supplementary Materials*).

**Data availability:**

The original European XFEL data are located at DOI: 10.22003/XFEL.EU-DATA-008025-00 and will be publicly available in 2029 after the embargo period of 3 years.

For the purpose of open access, the author has applied a Creative Commons Attribution (CC BY) license to any Author Accepted Manuscript version arising from this submission.

**Acknowledgements**

We acknowledge European XFEL for the attribution of beamtime at the HED instrument under proposal number 8025. The authors are indebted to the HIBEF user consortium for the provision of instrumentation and staff that enabled this experiment.

We acknowledge DESY (Hamburg, Germany), a member of the Helmholtz Association HGF, for the provision of experimental facilities. Parts of this research were carried out at P02.2. Data was collected using Laser Heating Experiment at P02.2 operated by DESY Photon Science. Beamtime was allocated for proposal I-20241114.

E. E. and A .F. G. acknowledge support of the US National Science Foundation Grant EAR-2049127 and Carnegie Science. A.F.G. acknowledges support of the National Science Foundation Grant DMR-2200670. EE also acknowledged support from the DFG through SPP 2440 DeepDyn (grant SA2585/13-1). GM was supported by the ANR grant MIN-DIXI (ANR-22-CE49-0006). SMcW, NJ, ZY acknowledge support from ERC grant to SMvW. E.K. is funded by the European Union (ERC, HotCores, Grant No 101054994). Views and opinions expressed are however those of the author(s) only and do not necessarily reflect those of the European Union or the European Research Council. Neither the European Union nor the granting authority can be held responsible for them

This result is part of a project that has received funding from the European Research Council (ERC) under the European Union's Horizon 2020 research and innovation programme (Grant agreement No. 101002868).

This work was supported by grant nos. EP/S022155/1 and EP/Z533671/1 (M.I.M. and J.M.) from the UK Engineering and Physical Sciences Research Council.

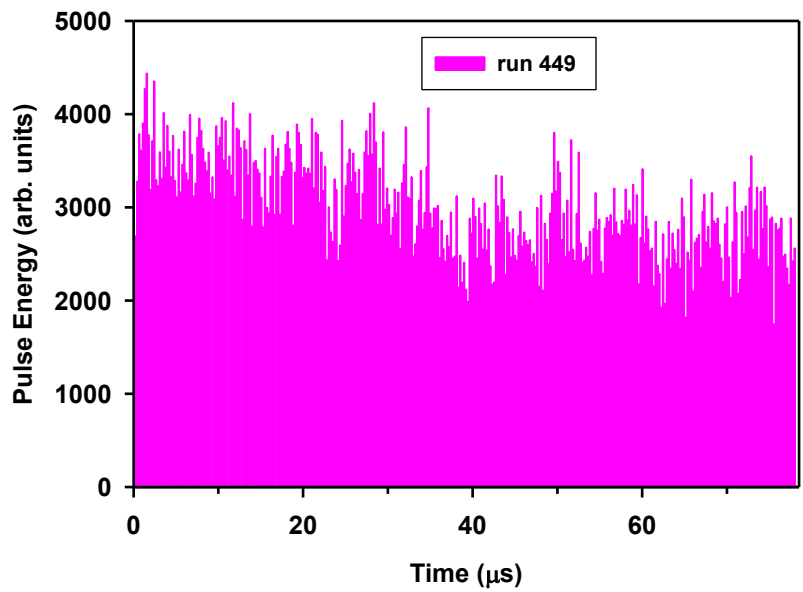

**Figure S1.** X-ray Free-Electron Laser (XFEL) pulse-to-pulse transmission measured using a photodiode.

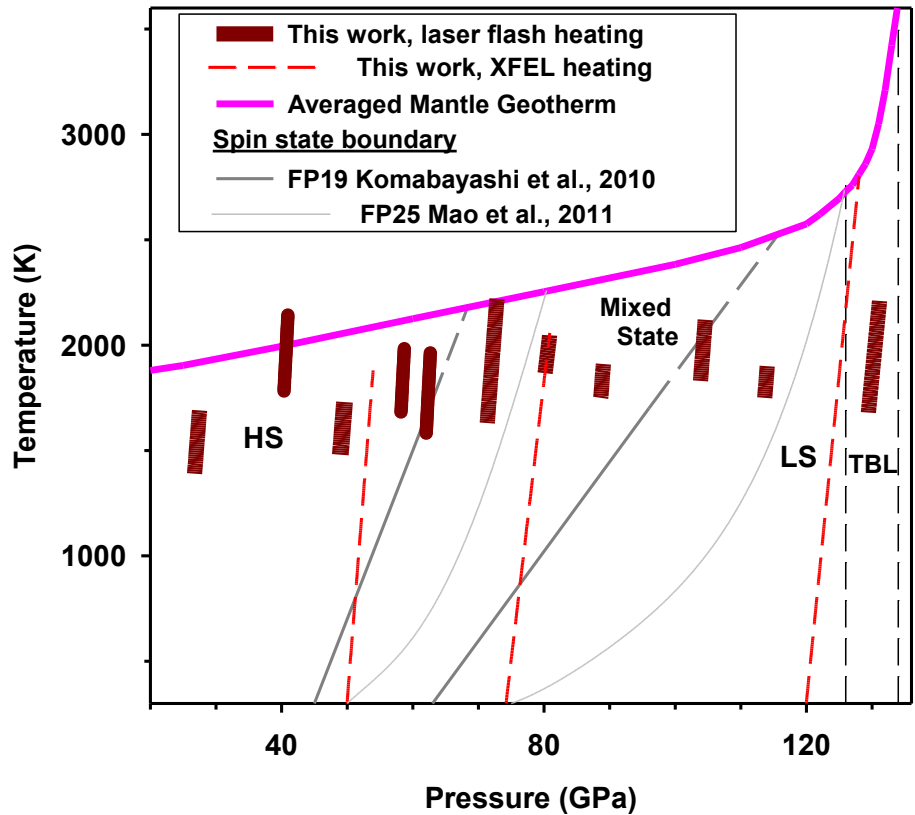

**Figure S2.** Pressure-temperature pathways in laser flash and XFEL heating experiments in a DAC performed in this work. The temperatures in flash heating (FH) experiments were determined by spectroradiometry, while in XFEL experiments the temperature was inferred from the thermal expansion of the sample measured by in situ synchrotron XRD technique using the thermal equation of state (EOS) of ferropericlase (FP) [1,2]. The thermal pressure was calculated based on an assumption of 30% value of the thermodynamic isochoric thermal pressure [3].

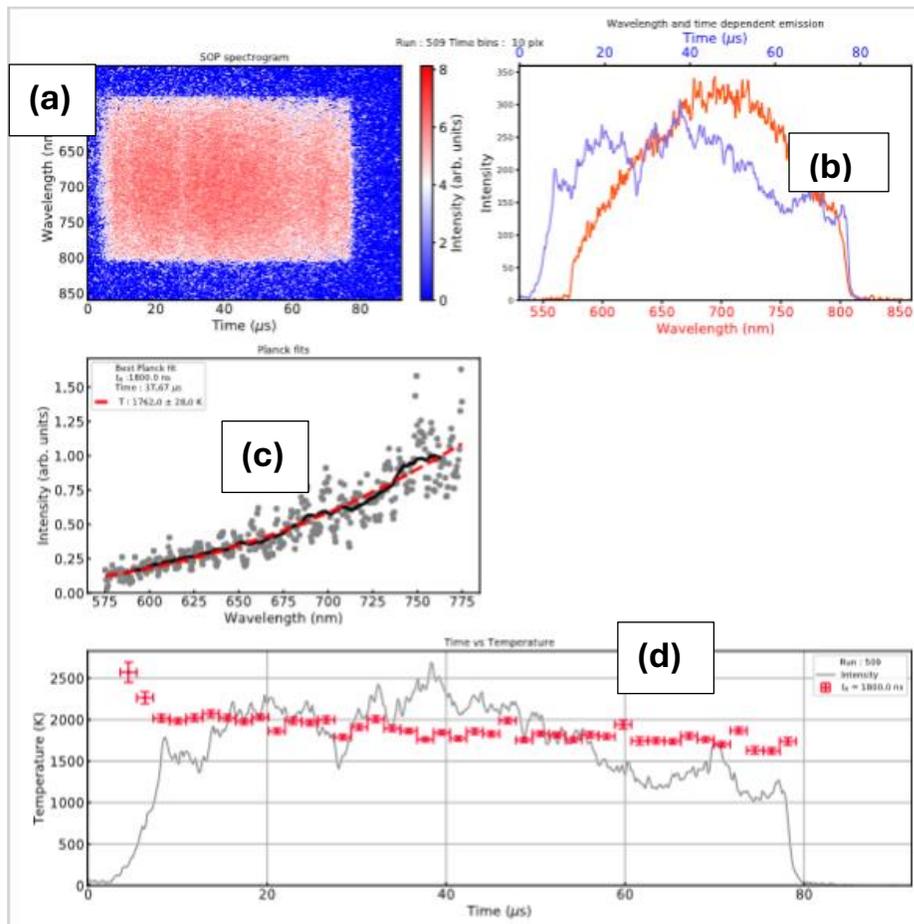
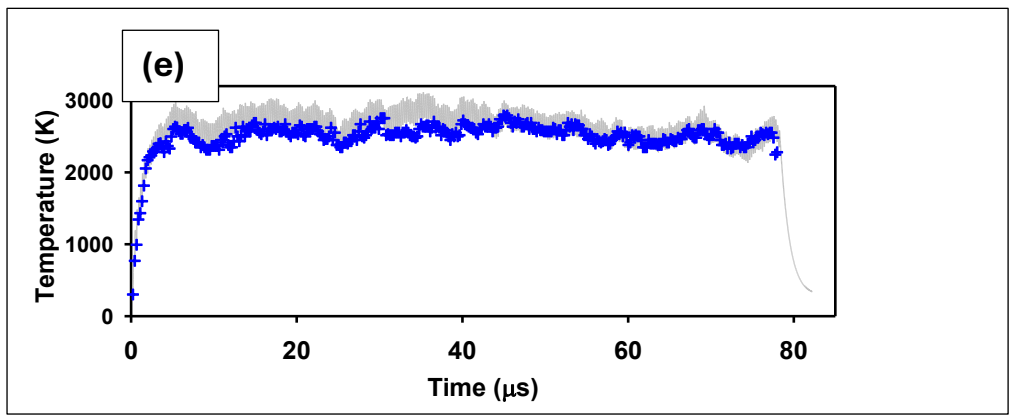

**Figure S3.** XFEL heating/XRD experiments of FP at a nominal pressure of 120 GPa at different times (run 509). (a) streak camera spectrogram; (b) integrated total white light intensity as a function of time and wavelength; (c) an example of the thermal radiation spectrum collected within the time window of 5 µs along with the Planck fit; (d) the measured temperature history superimposed with the intensity history; (e) heating history of FP determined from the thermal expansion data (blue crosses) compared to the best fitted Finite Element calculations (gray lines).

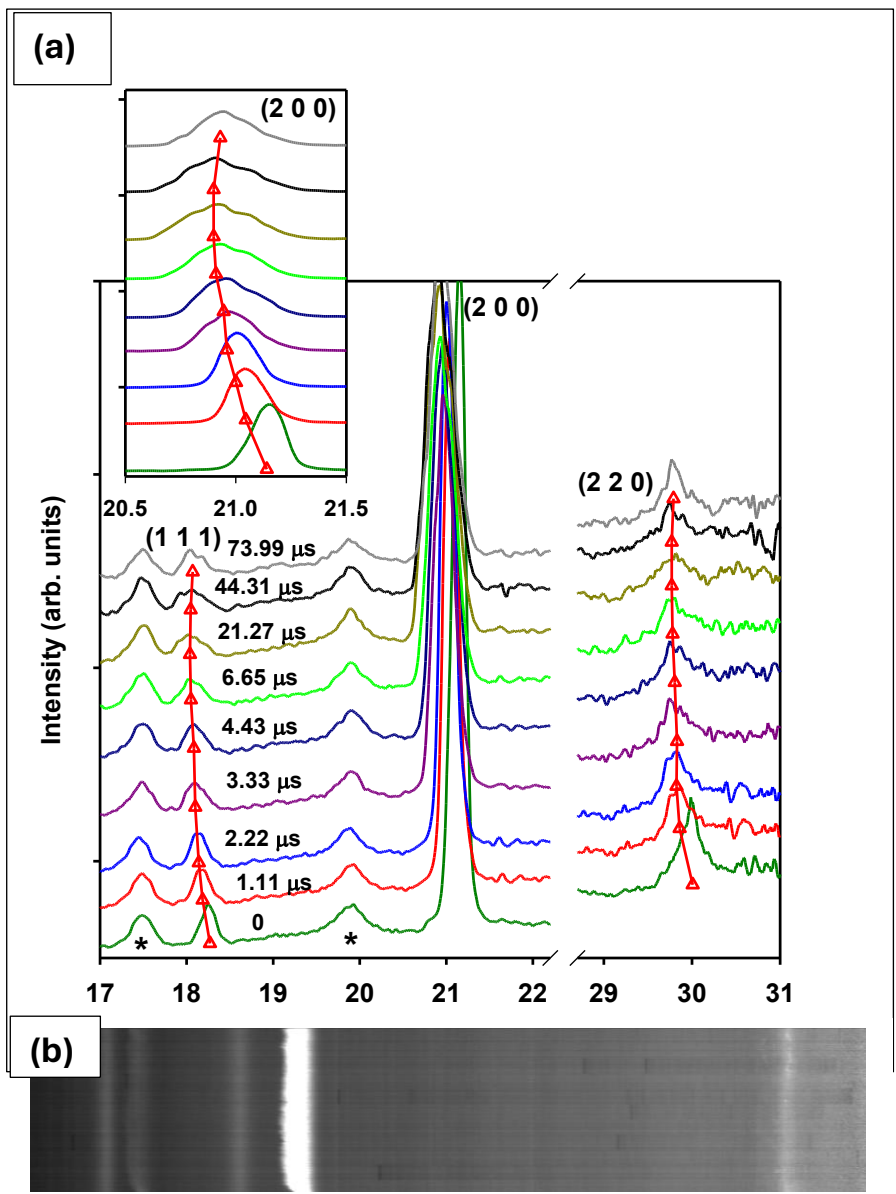

**Figure S4.** XFEL heating/XRD experiments. (a) Selected integrated 1D XRD patterns of FP at a nominal pressure of 120 GPa at different times (run 509). Open red triangles connected by lines illustrate the thermal shift of XRD peaks. The inset shows an extended view of the behavior of the (200) reflection. (b) 2D interferogram of the same data. Time is a vertical axis, and the diffraction angle is a horizontal axis.

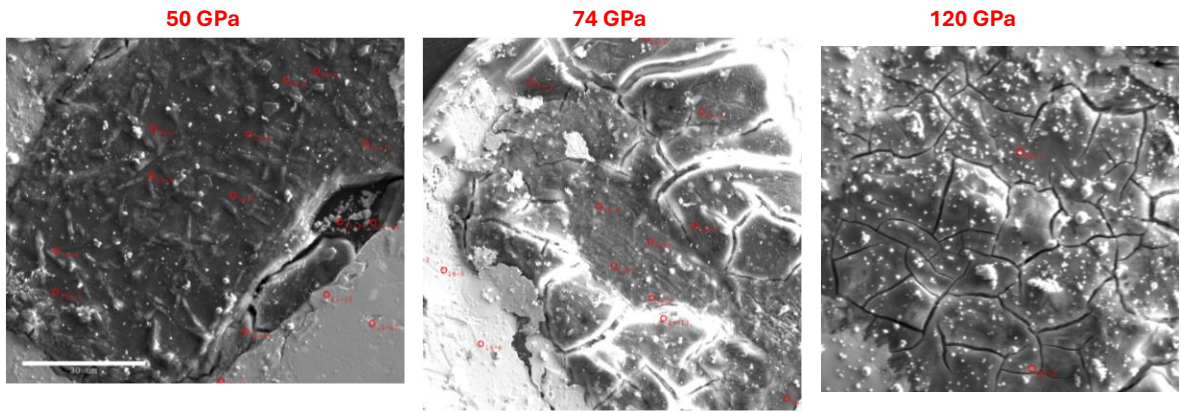

**Figure S5.** Scanning electron microscopy (SEM) images created by secondary electrons of the recovered to ambient FP samples after XFEL experiments at different pressures. The scale bar in the first image is 30 µm, the second image is around 100 µm wide, and the third image is around 40 µm wide.

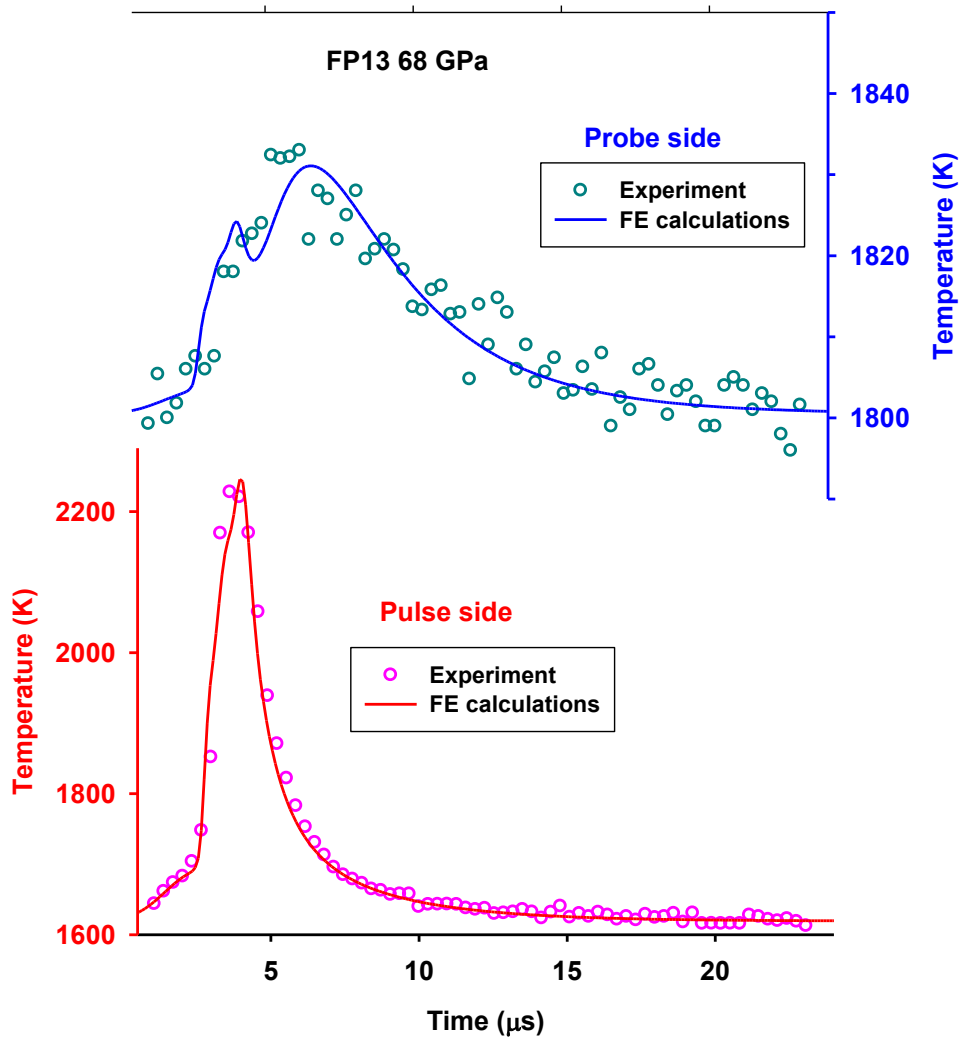

**Figure S6.** Heating histories in an *in-house* laser flash-heating experiment in FP at 68 GPa. Open circles (pink and green symbols correspond to the pulsed-heated and probe sample sides, respectively) are the experimentally determined radiative temperatures. The solid lines are the best fits from finite element (FE) calculations [4,5]. A narrow peak in the FE calculation curve on the probe side is due to an assumption of direct pulsed heating leaking to the opposite sample side (because of imperfections in the polarization optics). The parameters used in the fit are presented in Table S2.

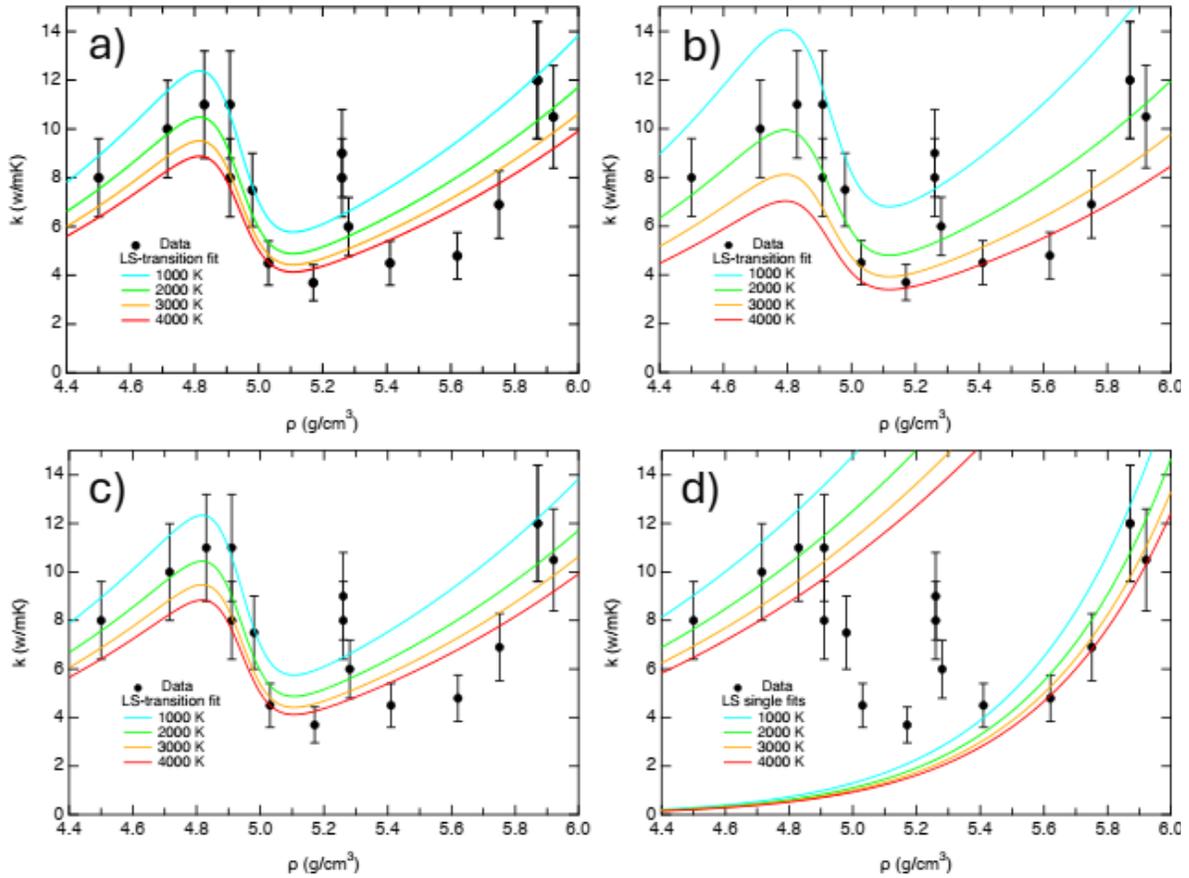

**Figure S7.** Liebfried Schloemann (LS) model [6] fits to thermal conductivity data at high pressure and temperature on FP. The models assume the high and low spin phases each follow the conventional LS formulation, $k = k_0(\rho/\rho_0)^g (T_0/T)^m$ and a sigmoidal transition of the form $k_h + (k_l - k_h)/\{1 + exp\,[-\kappa * (\rho - \rho_t)]\}$, where subscripts $h$ and $l$ refer to high- and low-spin phase parameters, and $\rho_t$ is the transition density with width defined by $\kappa$; units are W/mK, g/cm$^3$, and K for $k$, $\rho$, and $T$. Data, with 20% error bars (Fig. 3) are fitted using mean temperature in optical experiments and plateau temperature in XFEL experiments, assuming $\rho_0$=3.8 g/cm$^3$, $T_0$=300 K.

a) Fit assuming $m_h = m_l = 0.24$ (Table S1) and $g_h = g_l$, yielding $g_h = g_l = 5.80$, $k_{0h} = 4.46$, $k_{0l} = 1.31$, $\kappa = 20.7$ and $\rho_t = 4.93$.

b) Fit assuming $m_h = m_l = 0.5$ and $g_h = g_l$, yielding $g_h = g_l = 6.23$, $k_{0h} = 6.56$, $k_{0l} = 1.80$, $\kappa = 17.2$ and $\rho_t = 4.92$.

c) Fit assuming $m_h = m_l = 0.24$, yielding $g_h = 5.56$, $g_l = 5.79$, $k_{0h} = 4.66$, $k_{0l} = 1.31$, $\kappa = 21.2$ and $\rho_t = 4.93$. This is indistinguishable from (a).

The transition-including fits suggest a spin transition in the range of 45-60 GPa.

d) Here we made separate LS fits below (<4.9 g/cm$^3$,<50 GPa) and above (>5.3 g/cm$^3$, >75 GPa) the possible transition region with $m$=0.24 assumed. For high spin, $k_{0h} = 5.48$ and $g_h = 4.67$ and for low spin, $k_{0l} = 0.035$ and $g_l = 14.2$.

**Table S1. XFEL experiments in FP. Parameters that were used in FE calculations.**
Density and heat capacity are temperature-dependent following Refs. [1,7].

| Run # | P at 300 K (GPa) | Fe/(Fe+Mg) | XRD T range (K) | Sample thickness (μm) | k (W/m/K) (const) | k(W/m/K) T dependence | k(W/m/K) at 2000 K |
|---|---|---|---|---|---|---|---|
| 322 | 51 | 0.091(5) | 300-1130 | 9.7 | 11 | $15*(300/T)^{0.24}$ | 9.5-11 |
| 323 | 51 | 0.091(5) | 300-1880 | 9.7 | 8 | $7*(1+0.0003*(T-300))$ | 8-10.7 |
| 449 | 75 | 0.12(1) | 300-1600 | 9.6 | 9 | $18*(300/T)^{0.5}$ | 7-9 |
| 450 | 75 | 0.12(1) | 300-2050 | 9.6 | 8 | $12*(300/T)^{0.33}$ | 6.4-8 |
| 506 | 120 | 0.13(1) | 300-2100 | 7.1 | 12 | $15*(300/T)^{0.24}$ | 11.4-12 |
| 509 | 120 | 0.13(1) | 300-2800 | 7.1 | 12 | $18*(300/T)^{0.24}$ | 11.4-12 |

**Table S2. Flash heating experiments in FP. Parameters that were used in FE calculations.**

| Run# | P at 300 K (GPa) | Optical T range (K) | $k_{FP}$ (W/m/K) | $\rho_{FP}$ (g/cm³) | $C_P$ (J/kg/K) | FP thickness (μm) | $k_{KCl}$ (W/m/K) |
|---|---|---|---|---|---|---|---|
| 5.1 | 24 | 1390-1650 | 8 | 4.5 | 1200 | 5.16 | 16 |
| 6.1 | 36 | 1780-2140 | 10 | 4.715 | 1150 | 6.57 | 20 |
| 6.2 | 46 | 1470-1720 | 11 | 4.83 | 1150 | 5.05 | 20 |
| 6.3 | 56 | 1680-2230 | 7.5 | 4.98 | 1270 | 4.03 | 16 |
| 7.1 | 60 | 1580-1960 | 4.5 | 5.03 | 1450 | 4.31 | 16 |
| 8.1 | 68 | 1620-2240 | 3.7 | 5.17 | 1450 | 2.92 | 12 |
| 7.2 | 75 | 1860-2060 | 6 | 5.28 | 1450 | 3.87 | 20 |
| 8.2 | 85 | 1750-1910 | 4.5 | 5.41 | 1450 | 2.55 | 12 |
| 8.3 | 100 | 1830-2150 | 4.8 | 5.62 | 1350 | 2.54 | 10 |
| 10.1 | 110 | 1750-1910 | 6.9 | 5.75 | 1140 | 3.17 | 12 |
| 10.2 | 125 | 1670-2210 | 10.5 | 5.92 | 1139 | 2.86 | 12 |